\documentclass{ws-procs9x6-cpt22}
\begin{document}

\newcommand{\refeq}[1]{(\ref{#1})}
\def\etal {{\it et al.}}

\title{Lorentz-Violating Electrodynamics in Continuous Matter}

\author{M.M.\ Ferreira Jr., P.D.S.\ Silva, and  M.\ Schreck}

\address{Departamento de F\'{i}sica, Universidade Federal do Maranh\~{a}o,\\ Campus
	Universit\'{a}rio do Bacanga, S\~{a}o Lu\'is (MA), 65080-805, Brazil}

\begin{abstract}
We review the propagation of electromagnetic waves in continuous matter in the presence of Lorentz-violating terms. First, we briefly discuss classical electrodynamics with regard to optical properties of a dielectric medium exhibiting the Chiral Magnetic Effect. Such a medium can be modeled by Maxwell--Carroll--Field--Jackiw theory. Second, we describe the impact of CPT-odd terms of mass dimensions three and five, respectively, on electromagnetic propagation in continuous matter. Birefringence and absorption are analyzed in the scenarios investigated. Our findings provide new insights into the interplay between classical field theory and optical effects in matter.
\end{abstract}

\bodymatter

\section{Introduction}

The Standard-Model Extension (SME)\cite{Colladay} is a framework for studying Lorentz violation in all particle sectors of the Standard Model. Recently, the CPT-odd photon sector of the SME, represented by Maxwell--Carroll--Field--Jackiw (MCFJ) electrodynamics, was addressed in the context of condensed-matter systems subject to the Chiral Magnetic Effect (CME).\cite{Qiu} This effect, which arises due to an asymmetry in the behavior between right- and left-handed fermions, describes the generation of an electric current via an applied magnetic field. The material property responsible for this current in a dielectric substrate was taken into account at the classical level, revealing intriguing optical phenomena such as isotropic birefringence.\cite{Pedro1}

Incorporating field operators with mass dimensions greater than four complemented the minimal SME and provided its nonminimal version.\cite{Koste1} Several aspects of higher-derivative electrodynamics \textit{in vacuo} were discussed in the literature, including a derivation of the gauge-field propagator and an analysis of classical causality and unitarity at tree-level for different photon modes.\cite{Leticia1} More recently, the latter theory was adapted to the description of a continuous medium, followed by an investigation of the optical properties in this setting.\cite{Pedro2} In the current work, we review some aspects of electromagnetic propagation in continuous media in two scenarios: i) a dielectric medium endowed with a magnetic conductivity; ii) a dielectric medium affected by CPT-odd terms of mass dimensions three and five.

\section{Magnetic-conductivity effects in continuous matter}

We start from the Maxwell equations, supplemented by the usual constitutive relations,
\begin{equation}
{\bf{D}}=\epsilon {\bf{E}}, \quad {\bf{H}}=\mu^{-1} {\bf{B}}\,, \label{eq:proceedings-1}
\end{equation}
with $\epsilon$ and $\mu$ being the electric permittivity and magnetic permeability, respectively. Let us also consider an extension of Ohm's law,
\begin{equation}
J^{i}=\sigma^{B}_{ij} B^{j}\,, \label{eq:proceedings-2}
\end{equation}
where $\sigma^{B}_{ij}$ is a general magnetic-conductivity tensor, describing a material property of the medium. Using a plane-wave \textit{ansatz}, $({\bf{E}}, {\bf{B}}) \propto \exp[\mathrm{i} ({\bf{k}}\cdot {\bf{r}}-\omega t) ] $ the Maxwell equations together with Eqs.~\eqref{eq:proceedings-1} and \eqref{eq:proceedings-2} yield
\begin{subequations}
\begin{equation}
M_{ij}E^{j} = 0\,,\quad M_{ij}=k^{2} \delta_{ij} - k_{i}k_{j} - \mu \bar{\epsilon}_{ij}(\omega)\,, \label{eq:proceedings-3}
\end{equation}
with the effective electric permittivity given by
\begin{equation}
\bar{\epsilon}_{ij}(\omega)= \epsilon \delta_{ij} + \frac{\mathrm{i}}{\omega} \sigma^{B}_{ia}\epsilon_{abj}n_{b}\,. \label{eq:proceedings-4}
\end{equation}
\end{subequations}
The dispersion equations, obtained through $\mathrm{det}[M_{ij}]=0$ and ${\bf{k}}=\omega {\bf{n}}$, determine the refractive indices $|\mathbf{n}|\equiv n$ of the medium. For an isotropic conductivity, $\sigma_{ij}^{B}=\Sigma \delta_{ij}$, being the common case for the CME, they read
\begin{equation}
n_{\pm} = \sqrt{\mu\epsilon + \left(\frac{\mu\Sigma}{2\omega}\right)^{2}} \pm \frac{ \mu\Sigma}{2\omega}\,. \label{eq:proceedings-5}
\end{equation}
The associated propagating modes are described by left- and right-handed circularly polarized electric-field vectors, yielding birefringence measured in terms of the rotatory power
\begin{equation}
\delta = -\frac{\omega}{2}[\mathrm{Re}(n_{+})-\mathrm{Re}(n_{-}) ]\,. \label{eq:proceedings-6}
\end{equation}
Applying the latter to the isotropic scenario implies $\delta = - {\mu\Sigma/2}$. Since the refractive indices in Eq.~(\ref{eq:proceedings-5}) are real for all frequencies, dichroism does not occur in this case.

We can also examine an antisymmetric current, whose conductivity $\sigma^{B}_{ij}=\epsilon_{ijk}b_{k}$ is parameterized in terms of an intrinsic three-vector ${\bf{b}}$. With the angle $\theta$ between ${\bf{n}}$ and ${\bf{b}}$, the corresponding refractive index is
\begin{equation}
n=\sqrt{\mu\epsilon -\frac{\mu^{2}}{4\omega^{2}} b^{2} \cos^{2}\theta} + \mathrm{i}\frac{\mu}{2\omega}b\cos\theta\,, \label{eq:proceedings-7}
\end{equation}
where $b\equiv |\mathbf{b}|$. The imaginary part in Eq.~\eqref{eq:proceedings-7} acts like an absorption coefficient, which implies that a propagating wave is subject to attenuation. Thus, off-diagonal magnetic-conductivity terms ascribe a conducting behavior to a dielectric substrate. The literature\cite{Pedro1} can be consulted for additional details on these kinds of phenomena.

\section{MCFJ electrodynamics in continuous matter}

In the following, we dedicate ourselves to CPT-odd MCFJ electrodynamics as well as a dimension-five extension of the latter. In continuous matter, these types of theories are governed by
\begin{equation}
\mathcal{L}=-\frac{1}{4} G^{\mu\nu} F_{\mu\nu} - \frac{1}{4} \epsilon^{\beta\lambda\mu\nu} V_{\beta} A_{\lambda} F_{\mu\nu} - A_{\mu} J^{\mu}\,,\label{eq:proceedings-MCFJ-1}
\end{equation}
where $V^{\mu}=(V^{0}, {\bf{V}})$ is a Lorentz-violating background vector, $G^{\mu\nu}=(1/2)\chi^{\mu\nu\alpha\beta}F_{\alpha\beta}$ the field-strength tensor in matter, and $\chi^{\mu\nu\alpha\beta}$ the constitutive tensor. In this case, the electromagnetic dispersion equation is\cite{Pedro2}
\begin{subequations}
\begin{equation}
\label{eq:MCFJ-medium-dispersion-equation}
\overline{p}^{4} + \overline{p}^{2} \overline{V}^{2} - (\overline{p} \cdot \overline{V} )^{2} =0\,,
\end{equation}
with a redefined four-momentum and background vector
\begin{equation}
\label{eq:effective-four-vectors}
\overline{p}^{\mu}\equiv\left(\sqrt{{\epsilon}}\,\omega,\frac{\mathbf{k}}{\sqrt{\mu}}\right)\,,\quad \overline{V}^{\mu}\equiv\left(\sqrt{\mu}\,V^0,\frac{\mathbf{V}}{\sqrt{{\epsilon}}}\right)\,.
\end{equation}
\end{subequations}
For the purely timelike sector of the background, the refractive indices read
\begin{equation}
n_{\pm} = \sqrt{\mu\epsilon + \frac{\mu^{2}V_{0}^{2}}{4\omega^{2}}} \pm \frac{\mu V_{0}}{2\omega}\,, \label{eq:proceedings-MCFJ-2}
\end{equation}
with $V_{0}$ playing the same role as does the isotropic conductivity $\Sigma$ in Eq.~\eqref{eq:proceedings-5}. The modes are described by circularly polarized waves with the rotatory power of Eq.~\eqref{eq:proceedings-6} given by $\delta = - {\mu V_{0}/2}$. For the purely spacelike scenario, we distinguish between two special cases: i) a ${\bf{V}}$-perpendicular configuration with ${\bf{n}}\cdot {\bf{V}} = 0$ and ii) a ${\bf{V}}$-longitudinal configuration where ${\bf{n}}\cdot {\bf{V}}=\pm n |{\bf{V}}|$. For the first, the refractive indices can be cast into the form
\begin{equation}
n_{\pm} =\sqrt{\mu\epsilon + \frac{\mu {\bf{V}}^{2}}{2\epsilon \omega^{2}} (-1\pm 1) }\,. \label{eq:proceedings-MCFJ-3}
\end{equation}
The propagating modes are linearly polarized and have an additional longitudinal component. Birefringence is characterized by the phase shift $\Delta = (2\pi/\lambda_{0}) d (n_{+}-n_{-} ) $, with the vacuum wavelength $\lambda_{0}$ of light and the thickness $d$ of the medium. For the second, the refractive indices read
\begin{align}
n_{\pm}=\sqrt{\mu\epsilon \pm \frac{\mu |{\bf{V}}|}{\omega}}\,, \label{eq:proceedings-MCFJ-4}
\end{align}
where the associated propagating modes are circularly polarized. Thus, the rotatory power is
\begin{equation}
\frac{\Delta}{d}=\frac{2\pi}{\lambda_0}\sqrt{\mu \epsilon}\left(1-\sqrt{1-\frac{\mathbf{V}^{2}}{\epsilon^2\omega^{2}}}\right)\,.
\label{phase-shift2}
\end{equation}
Furthermore, in the frequency range $0<\omega < |{\bf{V}}|/\epsilon$, the refractive index $n_{-}$ of Eq.~\eqref{eq:proceedings-MCFJ-4} becomes purely imaginary, so that one of the two circularly polarized modes is absorbed. This behavior is suitably characterized by the dichroism coefficient $\delta_{\mathrm{d}} = -(\omega/2)[\mathrm{Im}(n_{+}) - \mathrm{Im}(n_{-})]$, in particular,
\begin{align}
\delta_{\mathrm{d}}&= \frac{\sqrt{\mu\epsilon}}{2} \omega \sqrt{-1 + \frac{ | {\bf{V}}|}{\omega \epsilon}}\,.
\label{dichroism-longitudinal-MCFJ}
\end{align}
Finally, we explore a higher-derivative dimension-five extension of MCFJ electrodynamics in continuous matter based on\cite{Pedro2}
\begin{equation}
\mathcal{L}= -\frac{1}{4} G^{\mu\nu} F_{\mu\nu} +\frac{1}{2} \epsilon^{\beta\lambda\mu\nu} U_{\beta} A_{\lambda} \square F_{\mu\nu} -A_{\mu}J^{\mu}\,,
\label{lagrangian3}
\end{equation}
with the Lorentz-violating background vector $U^{\mu} = (U^{0},{\bf{U}})$. This theory gives rise to the following dispersion equation:
\begin{subequations}
\begin{equation}
\label{eq:MCFJ-dim5-medium-dispersion-equation}
\overline{p}^4+4p^4\left[\overline{U}^2\overline{p}^2-(\overline{U}\cdot\overline{p})^2\right]=0\,,
\end{equation}
expressed in terms of Eq.~\eqref{eq:effective-four-vectors} and a properly redefined background,
\begin{equation}
\overline{U}^{\mu}\equiv\left(\sqrt{\mu}\,{U}^0,\frac{{\mathbf{U}}}{\sqrt{{\epsilon}}}\right)\,. \label{eq:effective-U-four-vector}
\end{equation}
\end{subequations}
To analyze the purely spacelike scenario, it is again convenient to distinguish between two cases: i) a ${\bf{U}}$-perpendicular configuration and ii) a ${\bf{U}}$-longitudinal configuration, defined in analogy to the above. For the first,
\begin{equation}
n_{0}=\sqrt{\mu\epsilon}, \quad n_{\pm}^{2}=1+ \frac{\epsilon}{8\mu\omega^{2} {\bf{U}}^{2}} \left[-1 \pm \sqrt{1+16\mu^{2}\omega^{2} {\bf{U}}^{2}\left(1-\frac{1}{\mu\epsilon}\right)}\, \right]\,. \label{eq:proceedings-MCFJ-6}
\end{equation}
The refractive index $n_0$ corresponds the usual one of Maxwell's electrodynamics, while the remaining two, $n_{\pm}$, capture effects stemming from the dimension-five term in Eq.~\eqref{lagrangian3}. The refractive index $n_{+}$ is real for the whole frequency range and exhibits anomalous dispersion, whereas $n_{-}$ has a simple root, $\omega_{-}= \epsilon/(2|{\bf{U}}|)$. For frequencies below the latter critical value, $n_{-}$ becomes purely imaginary, which disqualifies this mode as a propagating one. The refractive index $n_{0}$ is accompanied by a linearly polarized electric field. Linear polarizations with additional longitudinal components are found for~$n_{\pm}$.

For the second configuration of $\mathbf{U}$ we obtain
\begin{equation}
n_{\pm}^{2}=\frac{\mu(\epsilon \pm 2\omega |{\bf{U}}|)}{1\pm 2\mu\omega |{\bf{U}}| }\,. \label{eq:proceedings-MCFJ-8}
\end{equation}
The mode associated with $n_{+}$ propagates for any frequency and exhibits anomalous dispersion. On the order hand, the behavior of the mode related to $n_{-}$ is more intricate and one finds distinct behaviors in three different regimes: i) for $0<\omega < \omega_{0}$, with $\omega_{0}=1/(2\mu|{\bf{U}}|)$, the mode propagates and $n_{-}$ increases very fast; ii) for $\omega_{0} <\omega < \omega_{-}$, with $\omega_{-}= \epsilon/(2|{\bf{U}}|)$, the refractive index $n_{-}$ becomes purely imaginary, whereupon propagation is prohibited; iii) for $\omega > \omega_{-}$, propagation is enabled again. As the propagating waves are circularly polarized, we evaluate their rotatory power:
\begin{equation}
\delta = -\frac{\sqrt{\mu\epsilon}\omega}{2}  \left( \frac{g_{+}}{\sqrt{S_+}} - \frac{ g_{-}}{\sqrt{S_-}} \right)\,, \quad g_{\pm} = \sqrt{1\pm \frac{2\omega}{\epsilon} |{\bf{U}}|}\,, \label{eq:proceedings-MCFJ-10}
\end{equation}
where $S_{\pm}=1 \pm 2\mu\omega |{\bf{U}}|$. In total, there are three physical modes in the dimension-five MCFJ theory defined in Eq.~\eqref{lagrangian3}, since an additional mode originates from the higher-derivative structure of the theory. This is the reason why we find the three physical refractive indices of Eq.~\eqref{eq:proceedings-MCFJ-6}. It is crucial to emphasize that the occurrence of more than two modes is not indicative of a loss of U(1) gauge invariance.\cite{Pedro2}

\section*{Acknowledgments}

The authors are grateful to the Brazilian research agencies CNPq, CAPES, and FAPEMA for invaluable financial support.

\end{document}